\documentclass[twocolumn]{aastex701}
\usepackage{amsmath}
\usepackage{multirow}

\newcommand{\oiii}{[O\,{\footnotesize III}]\,$\lambda5008$}

\newcommand{\kms}{$\,\mathrm{km\,s^{-1}}$}

\begin{document}

\title{JADES: Dynamical Measurements of Dark Matter Halo Masses at $z \approx 7$ using Galaxy Pairs}

\author[0000-0002-8876-5248]{Zihao Wu}
\affiliation{Center for Astrophysics $|$ Harvard \& Smithsonian, 60 Garden St., Cambridge MA 02138 USA}
\email{zihao.wu@cfa.harvard.edu}
\author[0000-0002-2929-3121]{Daniel J.\ Eisenstein}
\affiliation{Center for Astrophysics $|$ Harvard \& Smithsonian, 60 Garden St., Cambridge MA 02138 USA}
\email{deisenstein@cfa.harvard.edu} 
\author[orcid=0000-0002-9280-7594]{Benjamin D.\ Johnson}
\affiliation{Center for Astrophysics $|$ Harvard \& Smithsonian, 60 Garden St., Cambridge MA 02138 USA}
\email{benjamin.johnson@cfa.harvard.edu}  
\author[0000-0001-6950-1629]{Lars Hernquist}
\affiliation{Center for Astrophysics $|$ Harvard \& Smithsonian, 60 Garden St., Cambridge MA 02138 USA}
\email{lhernquist@cfa.harvard.edu}
\author[orcid=0000-0002-8224-4505]{Sandro Tacchella}
\affiliation{Kavli Institute for Cosmology, University of Cambridge, Madingley Road, Cambridge, CB3 0HA, UK}
\affiliation{Cavendish Laboratory, University of Cambridge, 19 JJ Thomson Avenue, Cambridge, CB3 0HE, UK}
\email{st578@cam.ac.uk}  
\author[orcid=0000-0002-4271-0364]{Brant E. Robertson}
\affiliation{Department of Astronomy and Astrophysics, University of California, Santa Cruz, 1156 High Street, Santa Cruz, CA 95064, USA}
\email{brant@ucsc.edu}  
\author[0000-0001-7997-1640]{Santiago Arribas}
\affiliation{Centro de Astrobiolog\'ia (CAB), CSIC--INTA, Cra. de Ajalvir Km.~4, 28850 - Torrej\'on de Ardoz, Madrid, Spain}
\email{arribas@cab.inta-csic.es}
\author[orcid=0000-0002-8651-9879]{Andrew J.\ Bunker}
\affiliation{Department of Physics, University of Oxford, Denys Wilkinson Building, Keble Road, Oxford OX1 3RH, UK}
\email{andy.bunker@physics.ox.ac.uk} 
\author[0009-0009-8105-4564]{Qiao Duan}
\affiliation{Kavli Institute for Cosmology, University of Cambridge, Madingley Road, Cambridge, CB3 0HA, UK}
\affiliation{Cavendish Laboratory, University of Cambridge, 19 JJ Thomson Avenue, Cambridge, CB3 0HE, UK}
\email{qd231@cam.ac.uk}
\author[orcid=0000-0003-4337-6211]{Jakob M. Helton}
\affiliation{Department of Astronomy \& Astrophysics, The Pennsylvania State University, University Park, PA 16802, USA}
\email{jakobhelton@psu.edu} 
\author[orcid=0000-0001-7673-2257]{Zhiyuan Ji}
\affiliation{Steward Observatory, University of Arizona, 933 N. Cherry Avenue, Tucson, AZ 85721, USA}
\email{zhiyuanji@arizona.edu} 
\author[0000-0001-8630-2031]{D{\'a}vid Pusk{\'a}s}
\affiliation{Kavli Institute for Cosmology, University of Cambridge, Madingley Road, Cambridge, CB3 0HA, UK}
\affiliation{Cavendish Laboratory, University of Cambridge, 19 JJ Thomson Avenue, Cambridge, CB3 0HE, UK}
\email{dp670@cam.ac.uk}
\author[orcid=0000-0002-5104-8245]{Pierluigi Rinaldi}
\affiliation{Department of Astronomy, The University of Texas at Austin, Austin, TX 78712, USA}
\affiliation{Cosmic Frontier Center, The University of Texas at Austin, Austin, TX 78712, USA}
\email{prinaldi@utexas.edu} 
\author[orcid=0000-0002-4622-6617]{Fengwu Sun}
\affiliation{Department of Astronomy, School of Science, Westlake University, Hangzhou, Zhejiang 310030, P.\ R.\ China}
\affiliation{Center for Astrophysics $|$ Harvard \& Smithsonian, 60 Garden St., Cambridge MA 02138 USA}
\email{sunfengwu@westlake.edu.cn} 
\author[orcid=0000-0003-4891-0794]{Hannah \"Ubler}
\affiliation{Max-Planck-Institut f\"ur extraterrestrische Physik (MPE), Gie{\ss}enbachstra{\ss}e 1, 85748 Garching, Germany}
\email{hannah@mpe.mpg.de} 
\author[0000-0003-3307-7525]{Yongda Zhu}
\affiliation{Steward Observatory, University of Arizona, 933 N. Cherry Avenue, Tucson, AZ 85721, USA}
\email{yongdaz@arizona.edu}

\begin{abstract}
We present dynamical measurements of dark matter halo masses at $z\approx7$ using galaxy pairs. Galaxy pairs in the early Universe are usually in their first infall, before substantial orbital evolution or phase mixing. Their relative velocities closely trace the halo mass, especially when the pair separation is near the halo virial radius. In the TNG100 simulation, we find that dynamical mass estimators can recover halo masses with an intrinsic scatter of $\sim$0.2~dex. We apply this method to 10 pair systems at $z=6$--$8$ with \oiii\ spectroscopy from the JWST Advanced Deep Extragalactic Survey (JADES). We infer the stellar-to-halo mass relation while marginalizing over projection effects, measurement uncertainties, and intrinsic scatter. We find a mean halo mass of $\log\,(M_{200}/M_\odot)=11.06\pm0.21$ at $\log\,(M_{\star}/M_\odot)=9$. These galaxies do not appear to inhabit unusually massive halos. Our results instead indicate a high stellar-to-halo mass ratio, corresponding to an integrated star formation efficiency of $5^{+3}_{-2}\%$, about twice the TNG100 prediction, although the current statistical significance is limited.  Future observations with larger samples will tighten these constraints and directly examine whether enhanced star formation efficiency drives the overabundance of luminous galaxies at cosmic dawn.
\end{abstract}


\section{Introduction}

JWST has opened a new window onto the previously uncharted era of cosmic dawn. Observations are revealing galaxies within the first billion years that appear unexpectedly abundant, luminous, and massive \citep[e.g.,][]{Curtis-Lake2023NatAs, Carniani2024, Robertson2024, Naidu2025MoMz14}, which challenges galaxy formation theories and even cosmology \citep[e.g.,][]{BoylanKolchin2023,Shen2026}.

Measuring halo mass is key to understanding these phenomena. The stellar-to-halo mass ratio provides a direct measure of the integrated star formation efficiency (SFE), quantifying how efficiently the baryons associated with a halo are converted into stars. The stellar-to-halo mass relation (SHMR) further encodes how the SFE varies with stellar mass \citep{Moster2013, Behroozi2019}. Such measurements can clarify whether the ``overmassive problem'' is caused by a high SFE \citep[e.g.,][]{Dekel2023} or a high abundance of massive halos \citep[e.g.,][]{Shen2026}.

Recent studies find enhanced SFE for the most massive galaxies at high redshifts by comparing their stellar masses with the maximum halo masses expected within the surveyed volume \citep[e.g.,][]{Xiao2024, Turner2025, Lapasia2026}. In some cases, the inferred SFEs are so extreme that they even challenge $\Lambda$CDM cosmology \citep{BoylanKolchin2023, Menci2024}.  Halo mass estimates for the broader galaxy population, based on abundance matching and galaxy clustering, also suggest an enhanced SFE at high redshifts, with the inferred SFE rising by a factor of $\sim4$ from $z\sim4$ to $z\sim7$ \citep{Harikane2016} and by up to 1 dex by $z\gtrsim10$ \citep{Shuntov2025, Paquereau2025}. However, other studies with different sample selections find little evolution over $z\sim6$--$10$ \citep[e.g.,][]{Stefanon2021}.

Nonetheless, all of these halo-mass inferences are indirect. They are based on galaxy abundances or clustering statistics rather than directly probing the gravitational potentials of the halos themselves. Their results are therefore sensitive to sample selection, model calibration, and assumptions about the galaxy-halo connection \citep[e.g.,][]{Wechsler2018}. Several assumptions underlying these methods may break down in the early Universe. Stochastic star formation, for example, introduces substantial scatter into the relation between halo mass and galaxy luminosity or stellar mass, weakening the monotonic mapping assumed by abundance matching \citep[e.g.,][]{Mason2023, Mirocha2023, Sun2023}. Selection effects from low-redshift interlopers, post-burst dimming, and dust attenuation (e.g., \citealt{Endsley2025Burstiness, Simmonds2025, CurtisLake2026, Sun2026NIRCamDark}) may also bias halo-mass inferences based on galaxy statistics.

Dynamical methods provide a more direct measure of halo mass and are largely insensitive to these effects. Historically, rotation curves of spiral galaxies and velocity dispersions of galaxy clusters provided critical evidence for the existence of dark matter \citep{Zwicky1933, Rubin1970, Rubin1980}. Satellite kinematics have long provided halo-mass measurements for galaxies \citep{Zaritsky1994, More2011} and have been applied out to $z\sim1$ \citep{Conroy2007}.  At high redshifts, dynamical measurements have been applied to protocluster cores using their bright member galaxies \citep{Arribas2024}. However, measuring satellite kinematics at high redshifts is challenging because the satellites of most galaxies are too faint to identify. Gas kinematics have been used to constrain dark matter in galaxies out to $z\sim7$ \citep[e.g.,][]{Graaff2024, Fei2025, Danhaive2026}, but they measure only the mass within the innermost region of the halo, not the total halo mass. 

Gravitational lensing also probes halo mass directly and provides powerful measurements at lower redshifts \citep{Mandelbaum2006, Umetsu2020, Vegetti2024}. Suitable source--lens alignments, however, are exceedingly rare for galaxies acting as lenses at $z>7$. Lensing of the cosmic microwave background avoids the need for a background galaxy \citep{Madhavacheril2015}, but its angular resolution is currently insufficient to resolve the halos of typical galaxies at this redshift.

This work introduces a dynamical method of measuring halo mass using galaxy pairs. Galaxy pairs are common at high redshifts. We demonstrate that their relative velocities closely trace halo mass using the TNG100 simulation. We then apply this method to galaxy pairs at $z\approx7$ in the JWST Advanced Deep Extragalactic Survey (JADES). In this work, we adopt a flat $\Lambda$CDM cosmology with parameters from the full-mission Planck measurements (\hspace{-3pt}\citealt{Planck2020CMB}).

\begin{figure*}
    \centering
    \includegraphics[width=0.98\linewidth]{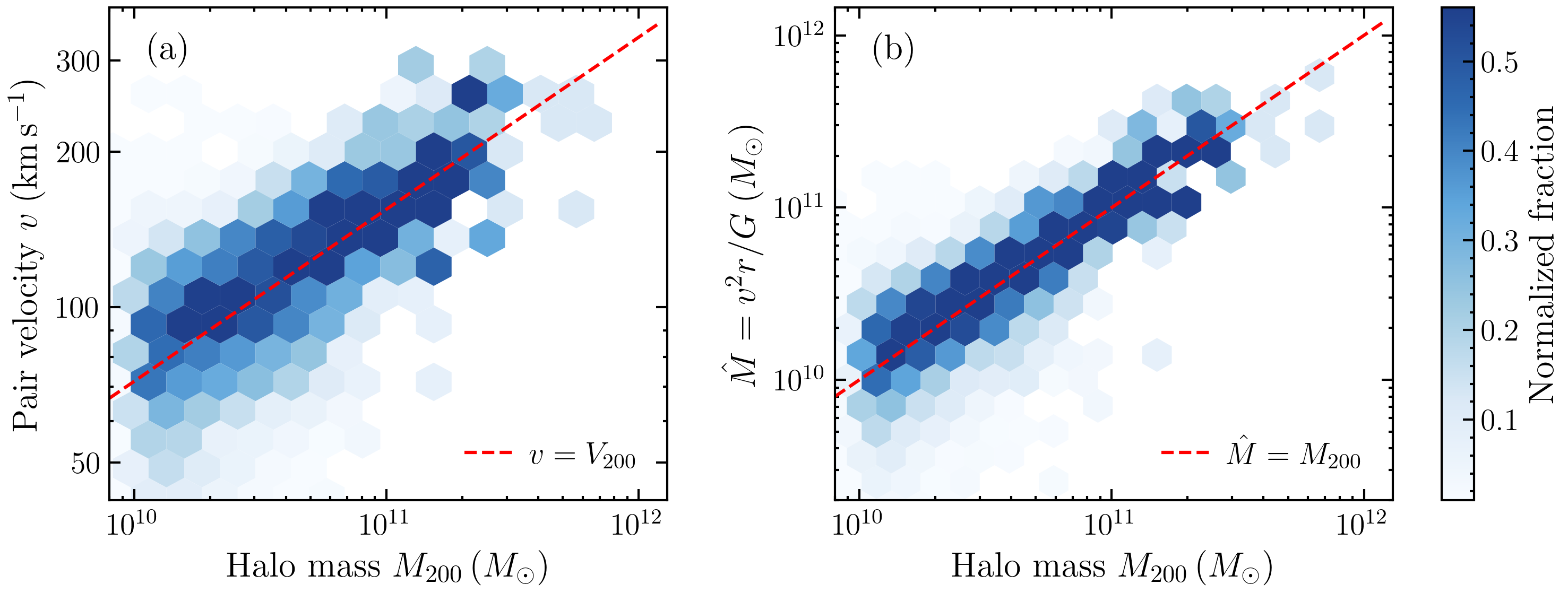}
    \caption{(a) Three-dimensional relative velocity $v$ of galaxy pairs versus halo mass $M_{200}$ in the TNG100 simulation at $z=7$. The red dashed line marks $v=V_{200}$ (Equation~\ref{eq:mass-vir}). (b) Dynamical mass estimate $\hat{M}=v^2r/G$ versus the true halo mass $M_{200}$, where $r$ is the three-dimensional pair separation. The red dashed line marks $\hat{M}=M_{200}$. In both panels,  hexbin counts are normalized independently within each halo-mass bin to show the relative distribution of pairs at fixed $M_{200}$. Only pairs with $0.6<r/R_{200}<1.5$ are included, where $R_{200}$ is the halo virial radius of the pair system. These panels show the intrinsic three-dimensional relations, while projection effects are modeled separately in Section~\ref{sec:projection}.}
    \label{fig:halo-mass-velocity}
\end{figure*}

\section{Method}
\label{sec:method}
We define the halo mass $M_{200}$ as the mass within $R_{200}$, the radius within which the mean density is 200 times the cosmic critical density \citep{Navarro1997}, and the halo virial velocity as $V_{200}\equiv\sqrt{GM_{200}/R_{200}}$, where $G$ is the gravitational constant. Because $R_{200}$ and  $M_{200}$ are connected by the cosmic critical density, $V_{200}$ is uniquely determined by $M_{200}$ at a given redshift. At $z=7$, for example, a halo with $M_{200}=10^{11}\,M_\odot$ has $R_{200}=18$~kpc ($3\farcs4$) and $V_{200}=154$\kms.

We first motivate the dynamical mass estimators for galaxy pairs, and calibrate them with the TNG100 simulation. We then analyze the statistical properties of projection effects, describe how to combine a sample of pairs to infer the galaxy-halo relation, and finally validate our method with mock observations.

\subsection{Physical Motivation}
\label{sec:motivation}
A commonly used dynamical mass estimator is $v^2r/G$, which measures the mass enclosed within a radius $r$ traced by an object moving at a speed $v$. On dimensional grounds, it should also apply to galaxy pair dynamics, with $v$ the relative velocity and $r$ the separation. In the matter-dominated early Universe, the two-body orbit with zero total energy takes the simple form $r\propto(GM)^{1/3}t^{2/3}$, where $M$ is the total gravitating mass associated with the pair \citep{Mo2010book}. Since $v\sim r/t$, we then have $v^2r/G\sim M$, up to a factor of order unity that depends on the orbital phase and on the initial energy and angular momentum. We calibrate this factor and its scatter with simulations.

We argue that this calibration factor should be close to unity when the pair separation is $R_{200}$. In the early Universe, a pair does not fall in from infinity, but from a turnaround radius, where it decouples from the Hubble expansion, reaches its maximum separation, and begins to collapse. If angular momentum is negligible, the relative motion of the two halos follows the same equations as the spherical collapse model, in which the turnaround radius is $2R_{200}$ \citep{Mo2010book}. The release of gravitational potential energy then gives $v=\sqrt{GM/R_{200}}=V_{200}$ at $r=R_{200}$, and hence $v^2r/G=M$. 

At separations $r<R_{200}$, however, we must account for how the halo profile shapes the potential. A useful feature of dark matter halos is that their gravitational potential varies only slowly at large radii, given the logarithmic term in the halo potential \citep{Navarro1997}. The relative velocity therefore depends only weakly on radius, remaining close to $V_{200}$. The virial velocity in turn determines the halo mass through

\begin{equation}
    M_{200} = V_{200}^3/(10\,GH(z)),
    \label{eq:mass-vir}
\end{equation}

\noindent
where $H(z)$ is the Hubble parameter at redshift $z$. The relative velocity of a pair alone therefore provides an estimate of the halo mass.

The two estimators are expected to agree at $r=R_{200}$, but their assumptions hold in different regimes: $v^2r/G$ relies on the infall picture valid at $r\gtrsim R_{200}$, whereas the velocity-only estimator relies on the flat potential inside the halo, at $r\lesssim R_{200}$. Their order-unity scalings should hold in general, but the exact coefficients depend on the initial conditions, the distribution of orbits, and the dynamics of halo mergers. These estimators therefore need calibration with simulations.

\subsection{Calibration with TNG100}
\label{sec:TNG}
We examine pair dynamics in the TNG100 simulation. TNG100 is one of the flagship simulations of the IllustrisTNG suite \citep{Marinacci2018, Naiman2018, Nelson2018, Pillepich2018, Springel2018, Nelson2019}, which models galaxy formation with magnetohydrodynamics in a $\sim$100 cMpc box, with dark-matter and baryonic mass resolutions of $7.5\times10^6\,M_\odot$ and $1.4\times10^6\,M_\odot$, respectively. We select galaxy pairs at $z=7$ with separations of $0.6$--$1.5\,R_{200}$. At these separations, both galaxies of a pair are assigned to the same halo in the TNG100 catalog, and the catalog halo mass $M_{200}$ corresponds to the total halo mass of the pair. Throughout this work, $M_{200}$ therefore refers to the mass of the common host halo assigned to the pair. When applying the estimators to observations, we assume that the selected close pairs are analogues of these same-halo systems.

We compute the velocity of each galaxy as the mass-weighted mean velocity of its stellar particles, after correcting for the Hubble flow. We do not use the subhalo velocities, which are dominated by dark matter and show $\sim$10\% systematic offsets from the stellar velocities, likely reflecting tidal distortion and stripping of subhalos in the merger process. We consider three-dimensional velocities in this subsection, so that the calibration reflects the scatter intrinsic to pair dynamics, while observational projection effects are treated in Section~\ref{sec:projection}. We adopt the stellar mass enclosed within twice the stellar half-mass radius, and correct for resolution-dependent differences by calibrating the TNG100 stellar masses to TNG50 at fixed subhalo mass, following the rescaling procedure of \citet{Engler2021}. We consider the total stellar mass of each halo by summing over its pair members, and find that the resulting SHMR of pairs in TNG100 is consistent with that of individual galaxies.

We confirm our expectation that the pair velocity is close to $V_{200}$, as shown in Figure~\ref{fig:halo-mass-velocity}a. Near a halo mass of $10^{11}\,M_\odot$, the median deviation from $V_{200}$ is smaller than 0.01 dex. The logarithmic scatter in the velocity is only 0.12 dex, which implies a precision of 0.36 dex in the halo mass estimate, given the relation in Equation~(\ref{eq:mass-vir}).

Figure~\ref{fig:halo-mass-velocity}b shows that the dynamical-mass estimator $\hat{M}=v^2r/G$ also recovers the halo mass accurately. TNG100 suggests that no additional calibration factor is needed: $\hat{M}$ already reproduces $M_{200}$ with a bias below 0.03 dex, confirming our expectation that the factor is close to unity. The scatter is only 0.23 dex, smaller than that of the velocity-only estimator.

\begin{figure}
    \centering
    \includegraphics[width=0.98\linewidth]{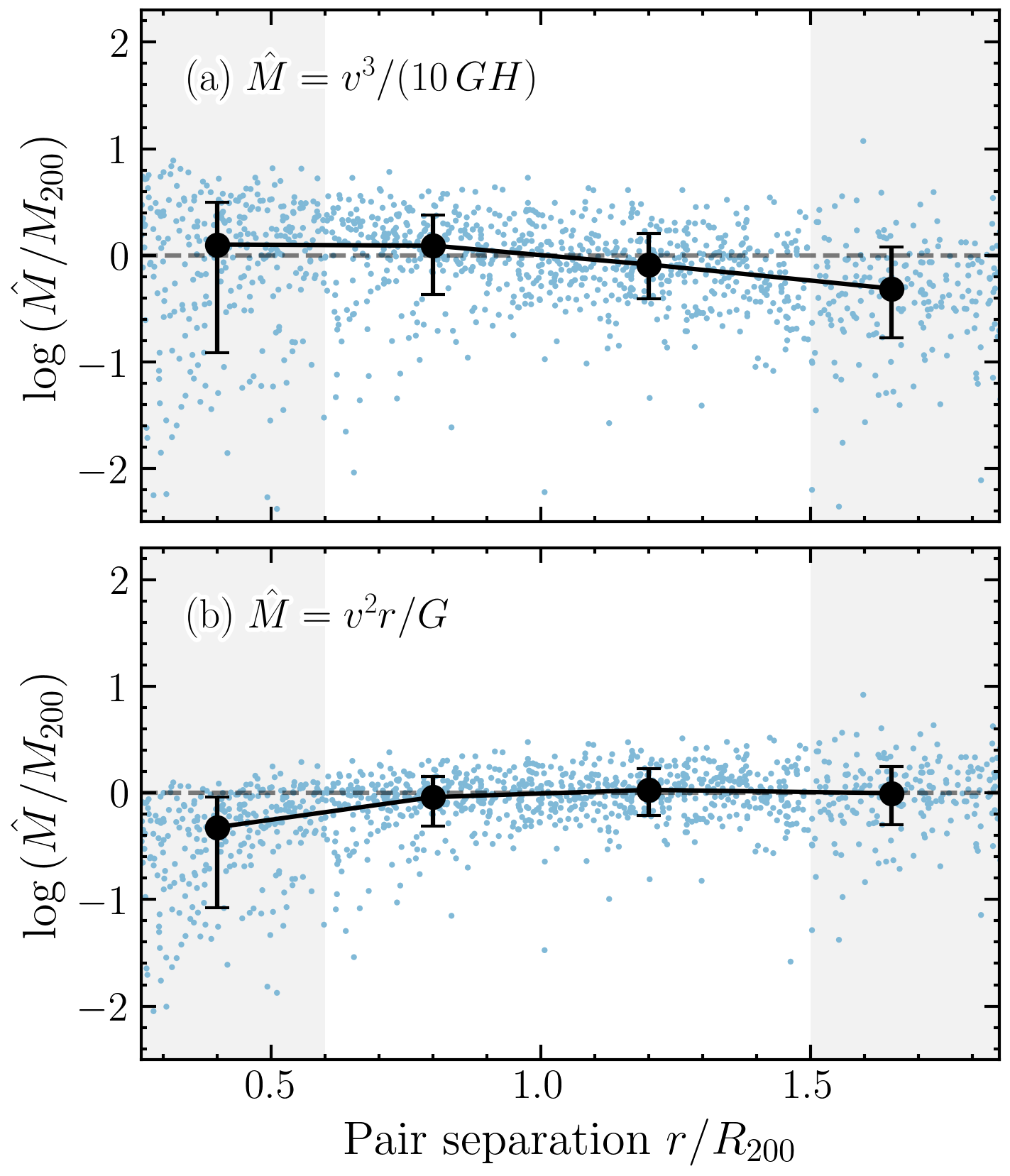}
    \caption{Accuracy of the halo mass estimate as a function of pair separation, for (a) the velocity-only estimator $\hat{M}=v^3/(10\,GH)$ and (b) the dynamical-mass estimator $\hat{M}=v^2r/G$ in TNG100 at $z=7$. Blue points show individual pairs, and black points with error bars show their binned medians and 16th--84th percentile ranges. The horizontal dashed line marks  $\hat{M}=M_{200}$. Gray shading indicates separations outside our selection window of $0.6$--$1.5\,R_{200}$. Line-of-sight projection effects are not included here.}
    \label{fig:vratio}
\end{figure}

Figure~\ref{fig:vratio} shows how the accuracy of the halo-mass estimate depends on the pair separation. As expected, both estimators are accurate at a separation of $R_{200}$. The velocity-only estimator performs best at $r\lesssim R_{200}$, but underestimates the halo mass beyond $R_{200}$: the bias is only 0.08~dex at $1.2\,R_{200}$, but grows to 0.3~dex at $1.7\,R_{200}$. At smaller separations, its bias stays below 0.1~dex even inside $0.5\,R_{200}$, although the scatter increases, likely due to dynamical friction \citep{Binney2008}. The $v^2r/G$ estimator performs well at $r\gtrsim R_{200}$ and starts to underestimate the halo mass inside $0.5\,R_{200}$. Over our selection window of $0.6$--$1.5\,R_{200}$, both estimators have a mean bias below 0.05~dex.

We find tentative evidence that the accuracy of both estimators depends on the stellar mass ratio $q_\star$ of the pair. The bias is smaller than 0.05 dex for $q_\star<0.5$, but increases to 0.08 and 0.16 dex for the $v^2r/G$ and velocity-only estimators, respectively, when $q_\star>0.5$. However, TNG100 has only nine pairs with $q_\star>0.5$ and $\log\,(M_\star/M_\odot)>8.5$ in our separation window, so the bias could arise from statistical fluctuations. Indeed, the bias is smaller than 0.05 dex in the $z=6$ snapshot of TNG100. We therefore do not correct for this effect, but treat it as a potential systematic uncertainty.

Our estimators remain accurate from $z=3$ to $z=10$, with biases below 0.1 dex. The scatter is nearly unchanged between $z=6$ and $z=10$, but increases toward lower redshift, growing to 0.30 and 0.44 dex at $z=3$ for the $v^2r/G$ and velocity-only estimators, respectively. The larger scatter likely reflects the more diverse dynamical histories at lower redshift.

Finally, we examine galaxy multiplets in TNG100. We decompose each multiplet into pairs relative to the most massive galaxy, and average their $\log\hat{M}$ to obtain the halo mass of the system. This averaging improves the precision: the scatter decreases to 0.19 and 0.31 dex for triplets, and to 0.14 and 0.21 dex for five-galaxy multiplets, for the $v^2r/G$ and velocity-only estimators, respectively. Multiplets are also more robust to projection uncertainties than pairs, because their members provide multiple sight lines (Section~\ref{sec:projection}).

\subsection{Projection effects}
\label{sec:projection}
We next consider projection effects in observations. Let $\theta$ be the angle between the relative velocity and the line of sight, $v_{\rm los}$ the line-of-sight velocity, and $r_{\rm proj}$ the projected pair separation. The velocity-only estimator becomes

\begin{equation}
    \hat{M}_{\rm vel} = \frac{v_{\rm los}^3}{10\,GH(z)}\frac{1}{|\cos\theta|^3},
    \label{eq:mass-vir-los}
\end{equation}

\noindent
and the dynamical-mass estimator is

\begin{equation}
    \hat{M}_{\rm dyn} = \frac{v_{\rm los}^2 \,r_{\rm proj}}{G}\frac{1}{\cos^2\theta\,\sin\theta}.
    \label{eq:mass-dyn-los}
\end{equation}

\noindent
The second equation assumes that the relative velocity is directed along the pair separation $r$, so that $r_{\rm proj}=r\sin\theta$. This is a good approximation: in TNG100, the median Binney anisotropy parameter \citep{Binney2008} of galaxy pairs is $\beta\equiv1-v_\perp^2/(2v_r^2)=0.8$. A more rigorous approach, which we adopt in our halo mass inference, is to marginalize over $\beta$ rather than assume purely radial motion. We nonetheless present Equation~(\ref{eq:mass-dyn-los}) here because it provides valuable intuition, and the halo masses it gives differ by less than 0.05~dex from the rigorous approach.

The dynamical-mass estimator is less sensitive to projection effects than the velocity-only estimator, owing to the additional $\sin\theta$ dependence of $r_{\rm proj}$. When the line of sight is nearly along the velocity direction, for example, the observed velocity approaches its maximum while the projected separation approaches its minimum, so the two effects partly cancel in their product.

Galaxy pairs have no preferred orientation relative to the observer, so their orientations are distributed isotropically. For an isotropic distribution, it is easy to derive that the  projection factor $u\equiv|\cos\theta|\sim\mathcal{U}(0, 1)$; i.e., $v_{\rm los}$ is uniformly distributed between 0 and the three-dimensional velocity $v$. The posterior distribution of halo mass can be  obtained from Monte Carlo (MC) simulations that draw random orientations. 

The projection uncertainty is purely statistical, so it can be reduced by observing a sample of galaxy pairs or galaxy multiplets. Our MC simulations show that the projection uncertainty in $\log M$ decreases roughly as $1/N$ with the number of pairs $N$. This scaling originates from the asymptotic behavior of the inferred $\log u$, which is known in non-regular inference theory \citep{Smith1985}. We find that the projection uncertainty falls to 0.2~dex with just five pairs. Beyond that, the intrinsic scatter of the estimators and the uncertainty in the velocity measurements dominate, and the total uncertainty follows the usual $1/\sqrt{N}$ scaling.

\subsection{Inference of the galaxy-halo relation}
\label{sec:inference}

Projection effects necessitate multiple pairs, sampling different sight lines, to constrain the halo mass tightly. A galaxy multiplet already contains several such sight lines, so its halo mass can be measured to high precision. For galaxy pairs, we instead need a sample of systems, which constrains only their mean halo mass. To make the problem more general, we constrain the SHMR, which encodes both the mean halo mass and its variation with stellar mass.

We present a framework for inferring the SHMR. At low redshift, the SHMR is usually parameterized as a broken power law, but at $z=7$, only the low-mass branch is present in TNG100 over the halo mass range $10^{10}$--$5\times10^{11}\,M_\odot$. We therefore model the SHMR as a single power law, which in logarithmic form is

\begin{equation}
    \log (M_\star/M_{\star,0}) = \alpha\log\,(M_{200}/M_\mathrm{h, 0}),
\label{eq:halo-stellar-mass}
\end{equation}

\noindent
where $M_{\star,0}$ is a reference stellar mass, $\alpha$ is the logarithmic slope, and $M_\mathrm{h,0}$ is the halo mass at $M_{\star,0}$. This relation has only $\alpha$ and $M_\mathrm{h,0}$ as free parameters, since $M_{\star,0}$ can be chosen arbitrarily. The TNG100 simulation suggests $\alpha=1.4$ and $M_\mathrm{h,0}=10^{11}\,M_\odot$ at $M_{\star,0}=4.2\times10^{8}\,M_\odot$. For a small sample, $\alpha$ is usually poorly constrained. We therefore infer only $\log M_\mathrm{h,0}$ here, fixing $\alpha=1.4$ to the TNG100 value. 

We use MC simulations to obtain the posterior of $\log M_\mathrm{h,0}$. For each pair, we draw a random realization of projection angles, velocity anisotropy parameter, measurement uncertainties, and intrinsic scatter following Sections~\ref{sec:TNG} and \ref{sec:projection}. We then obtain an MC sample of the halo mass $\log\,\tilde{M}_{200}$ from Equation~(\ref{eq:mass-vir-los}) or (\ref{eq:mass-dyn-los}), and of the stellar mass $\tilde{M}_\star$ from the posterior distribution obtained by spectral energy distribution (SED) fitting, where tilde accents denote MC realizations. This yields an MC sample of $\log M_\mathrm{h,0}$,

\begin{equation}
\log\,\tilde{M}_\mathrm{h,0} = \log\,\tilde{M}_{200}- \frac{1}{\alpha}\log\,\left({\tilde{M}_\star}/{M_{\star,0}}\right)+\tilde{\epsilon}.
\label{eq:mc}
\end{equation}

\noindent
We have included the intrinsic scatter of our halo mass estimator in $\log\,\tilde{M}_{200}$, and the  scatter of the SHMR in $\tilde{\epsilon}$. We obtain the intrinsic scatter by comparing the masses inferred from Equation~(\ref{eq:mass-vir-los}) or (\ref{eq:mass-dyn-los}) with the true halo masses in TNG100. For each estimator, we use the full residual distribution in the MC simulation because it has more outliers than a Gaussian distribution. Marginalizing over this distribution also automatically accounts for the small intrinsic bias ($\lesssim0.05$~dex) of the estimators found in TNG100.  We marginalize over $\tilde{\epsilon}$ to account for the deviation of individual halos from the mean SHMR, which has a scatter of $\sim$0.14~dex in TNG100. Lastly, we obtain the joint posterior by multiplying the likelihood function inferred from each pair.

Our formalism has implicitly assumed a uniform prior on $\log\,M_\mathrm{h,0}$. Such a prior is commonly regarded as uninformative. One might instead consider a prior based on the halo mass function, which favors low-mass halos that are more abundant. However, as $M_\mathrm{h,0}$ is the normalization of the SHMR rather than the mass of an individual halo, it is not clear why lower mass should be favored. Even for individual halos, such a prior should be multiplied by the observational selection function, which favors massive halos that may host brighter galaxies. The two effects act in opposite directions and may leave the effective prior close to our log-uniform choice.

Our formalism can be generalized to infer the halo mass--galaxy luminosity relation by replacing stellar mass with luminosity in Equation~(\ref{eq:mc}). In TNG100, the rest-frame UV luminosities from \citet{Vogelsberger2020}, including dust attenuation, have a logarithmic slope with halo mass similar to that of stellar mass, albeit with larger scatter. 

\subsection{Validation with mock observations}

We validate our method with mock observations from TNG100. We first consider noise-free mocks that include only line-of-sight projection effects. We select pairs in TNG100 with $\log\,(M_{200}/M_\odot)=10.95$--$11.05$. In each of 1000 realizations, we draw 10 pairs with random sight lines and infer their mean $\log M_{200}$ following Section~\ref{sec:projection}, using MC simulations to marginalize over projection effects, the velocity anisotropy parameter $\beta$, and the intrinsic scatter of the estimators. Both estimators recover the true halo mass: the median of the inferred $\log M_{200}$ agrees with the truth to better than 0.01~dex, and the formal error is consistent with the scatter of the inferred $\log M_{200}$ to within 10\%. This test therefore validates our treatment of projection effects and intrinsic scatter.

We further include measurement errors and sample distributions. We assume a velocity uncertainty of 30\kms\ and a stellar-mass uncertainty of 0.2~dex, and sample 10 pairs uniformly in $\log\,(M_\star/M_\odot)$ between 8 and 10. We generate 1000 random realizations of the sample and sight lines, and infer the mean halo mass at $\log\,(M_\star/M_\odot)=9$ following Section~\ref{sec:inference}, fixing $\alpha=1.4$ in the SHMR. The halo mass is recovered with a bias below 0.05~dex, and the formal error again agrees with the scatter of the inferred means to within 10\%. We therefore consider our treatment of measurement errors and our inference of the SHMR to be reliable.

\section{Observations and Measurements}
\label{sec:sample}

\begin{deluxetable}{ccccccc}
\tabletypesize{\scriptsize}
\setlength{\tabcolsep}{5.5pt}
\tablecaption{Properties of the Galaxy Pair Sample
\label{tab:pair_sample}}
\tablehead{
\colhead{Label} &
\colhead{$z$} &
\colhead{Pair IDs} &
\colhead{$r_{\rm proj}$} &
\colhead{$v_{\rm los}$} &
\colhead{$\log(M_{\star, \rm tot}/M_\odot)$} &
\colhead{$M_{\rm UV, tot}$}\\
\colhead{} &
\colhead{} &
\colhead{} &
\colhead{(kpc)} &
\colhead{(\kms)} &
\colhead{}&
\colhead{}
}
\startdata
Pair1  & 6.0490 & 1003743--1003962 & 17.6 & $168 \pm 34$ & $8.49^{+0.28}_{-0.19}$ & $-20.11 \pm 0.10$
 \\
Pair2  & 6.1263 & 68395--68872     &  9.1 & $21 \pm 37$  & $8.61^{+0.10}_{-0.09}$ & $-19.73 \pm 0.08$
\\
Pair3  & 6.1650 & 32002--169896    & 11.7 & $51 \pm 15$  & $8.21^{+0.26}_{-0.21}$ & $-18.93 \pm 0.14$
\\
Pair4  & 6.7215 & 1006478--1027155 & 14.4 & $58 \pm 32$  & $9.28^{+0.15}_{-0.14}$ & $-20.26 \pm 0.11$
\\
\multirow{2}{*}{Pair5}
    & 6.7452 & 1015068--1031196 &  8.1 & $6 \pm 35$   & $9.59^{+0.18}_{-0.17}$ & $-20.91 \pm 0.04$
\\
    & 6.7469 & 1015068--1031111 & 23.4 & $138 \pm 33$ & $9.59^{+0.18}_{-0.17}$ & $-20.91 \pm 0.04$
\\
Pair6  & 6.7601 & 1010806--1010816 & 18.7 & $38 \pm 40$  & $9.89^{+0.11}_{-0.13}$ & $-20.12 \pm 0.12$
\\
Pair7  & 6.9061 & 187025--187193   & 14.3 & $54 \pm 35$  & $8.90^{+0.11}_{-0.13}$ & $-21.10 \pm 0.08$
\\
Pair8  & 7.0349 & 1000741--1000905 & 22.5 & $39 \pm 42$  & $9.28^{+0.12}_{-0.25}$ & $-20.50 \pm 0.10$
\\
Pair9  & 7.2415 & 154775--219057   & 21.8 & $62 \pm 35$  & $8.51^{+0.14}_{-0.09}$ & $-20.57 \pm 0.05$
\\
Pair10 & 7.8829 & 175729--175837   &  8.0 & $37 \pm 30$  & $8.68^{+0.31}_{-0.17}$ & $-19.56 \pm 0.17$
\\
\enddata

\tablecomments{
Pairs used for halo mass inference, labeled Pair1--Pair10 in order of increasing redshift.
The Pair IDs column gives the JADES NIRCam catalog IDs of the two member galaxies.
Pair5 is a triplet, for which we list the two pairs defined relative to its
most massive member (ID 1015068). $z$ is the mean redshift of the pair.
$r_{\rm proj}$ is the proper projected separation.
$v_{\rm los}$ is the line-of-sight velocity difference measured from the [O\,{\tiny III}]\,$\lambda5008$ line, including measurement and systematic uncertainties (Appendix~\ref{sec:wave_cal}).
$M_{\star, \rm tot}$ is the total stellar mass, and $M_{\rm UV, tot}$ is the rest-frame UV absolute magnitude obtained by summing the fluxes of all galaxies in the system.
}
\end{deluxetable}

We search for galaxy pairs in the JADES fields \citep{Rieke2020, Bunker2020, Eisenstein2025Jof, Eisenstein2026JADES}, which have extensive JWST/NIRSpec and NIRCam wide-field slitless spectroscopy (WFSS).
We use the NIRCam imaging and photometric catalog from JADES Data Release 5 \citep{Johnson2026, Robertson2026}, NIRSpec spectra from JADES Data Release 4 \citep{CurtisLake2026, scholtz_jades_2026}, and NIRCam/WFSS spectra from the CONGRESS and FRESCO surveys (\citealt{Oesch2023, Meyer2024, CoveloPaz2025, Lin2026ApJ}; F. Sun in prep.) and JADES program 4540 (PI: Eisenstein; \citealt{Sun2026_4540}). Details of the data reduction are provided in these references. The deep F090W imaging in JADES is especially valuable at $z\approx7$: even when relatively shallow spectroscopy detects only the \oiii\ line, the Ly$\alpha$ dropout  in F090W can secure its redshift.

We select galaxies from a compiled catalog of all spectroscopic redshifts available within the JADES footprints \citep{Puskas2025, Puskas2025a}, retaining those at $6<z<8$ with \oiii\ detected at a signal-to-noise ratio (SNR) above 4. We use \oiii\ because it is the strongest emission line accessible to NIRSpec and NIRCam/WFSS at $z>7$. This redshift range extends the measurement to the highest redshift permitted by the current data while retaining a viable sample of galaxy pairs. We find no suitable pairs at $z>8$ in the current spectroscopic data.

We select pairs with projected separations of $8$--$25$\,kpc ($1\farcs5$--$4\farcs7$ at $z=7$) and redshift differences of $\Delta z<0.02$ ($\sim$750\kms). The separation range roughly corresponds to the $0.6$--$1.5\,R_{200}$ regime identified in Section~\ref{sec:method} after considering projection effects. This mapping is not sensitive to halo
mass, because $R_{200}\propto M_{200}^{1/3}$. The projected separation is usually close to the true separation: for an isotropic distribution, the probability that the projected separation exceeds half of the true separation is $\sqrt{3}/2\simeq87\%$. The $\Delta z<0.02$ criterion removes chance projections without rejecting true pairs, whose velocities rarely reach such high values (Figure~\ref{fig:halo-mass-velocity}). Under this criterion, the probability of foreground contamination is only $\sim10^{-3}$ per galaxy, according to the galaxy luminosity function of \citet{Harikane2025clump}.

We further inspect each galaxy visually and exclude highly irregular and very extended sources, because their internal kinematics can bias the velocity measurement. This concern is most serious for NIRSpec, whose microshutter may sample only part of a galaxy. We therefore require NIRSpec targets to be visually compact. NIRCam/WFSS, by contrast, captures the entire emission and averages over internal kinematics, so we allow its targets to be slightly extended if their morphologies are relatively smooth.

Our selections yield 10 systems at $z= 6$--$8$, comprising nine pairs and one triplet, as listed in Table~\ref{tab:pair_sample}. The triplet contributes two pairs, each defined relative to its most massive member, so the final sample contains 11 pairs in total. For the triplet, we marginalize over the intrinsic scatter of its two pairs jointly, using the scatter calibrated for triplets in Section~\ref{sec:method}. Two pairs contain a third galaxy within 6 kpc of the most massive member, below our adopted separation range. These additional galaxies are excluded from the pair sample but included in the total stellar mass of their respective systems.

We measure the line-of-sight relative velocities from redshift offsets,  such that $v_\mathrm{los} = c\, \Delta z/(1+z)$, where $c$ is the speed of light, $\Delta z$ is the redshift difference between the two galaxies, and $z$ is their mean redshift. For a pair at  $z=7$,  a typical line-of-sight velocity of 70\kms\ corresponds to $\Delta z =0.002$.  We measure redshifts by fitting the \oiii\ line with a Gaussian profile. We integrate the profile over each spectral pixel to account for undersampling. We account for noise correlations between spectral channels in the fitting. Spectra from different instruments are fitted independently.

We next assess the measurement uncertainties from wavelength calibration systematics. Although JADES Data Release 4 improves substantially upon the STScI pipeline calibration using a new data-driven model \citep{scholtz_jades_2026}, we still find mild residual instrumental offsets. We evaluate these offsets by measuring redshifts with the same fitting pipeline for all galaxies in the JADES NIRSpec catalog observed with more than one instrument. We present the details in Appendix~\ref{sec:wave_cal}. According to the calibration results in Table~\ref{tab:systematics}, we correct for the systematic bias and propagate the systematic scatter into the error budget. When a galaxy is observed with multiple instruments, we adopt the redshift from the instrument with the smallest systematic scatter, which gives the order of preference NIRSpec/G395H, G235M, G395M, and NIRCam/WFSS. The redshifts from different instruments are usually consistent within $1\sigma$ after the calibration, as shown in Table~\ref{tab:detail}.

We adopt the stellar masses from the JADES Data Release 5 catalog \citep{Duan2026}. These masses were inferred with the \texttt{Prospector} framework \citep{Johnson2021ApJS}, fixing the redshifts to the spectroscopic values and assuming a \citet{Chabrier2003} initial mass function (IMF) and a non-parametric star formation history (SFH) based on an empirically motivated, star-forming main-sequence prior. \citet{Duan2026} find that alternative SFH priors change the stellar mass by $\sim$0.2~dex. As an independent check, we perform our own \texttt{Prospector} fits of these galaxies with a flat prior and with the rising prior of \citet{Wu2025}, and find no systematic offset larger than 0.15~dex. Nevertheless, we note that other effects, such as a potentially top-heavy IMF in the early Universe, may introduce additional systematic uncertainties \citep{Wang2024IMF}.

\begin{figure*}
    \centering
    \includegraphics[width=0.98\linewidth]{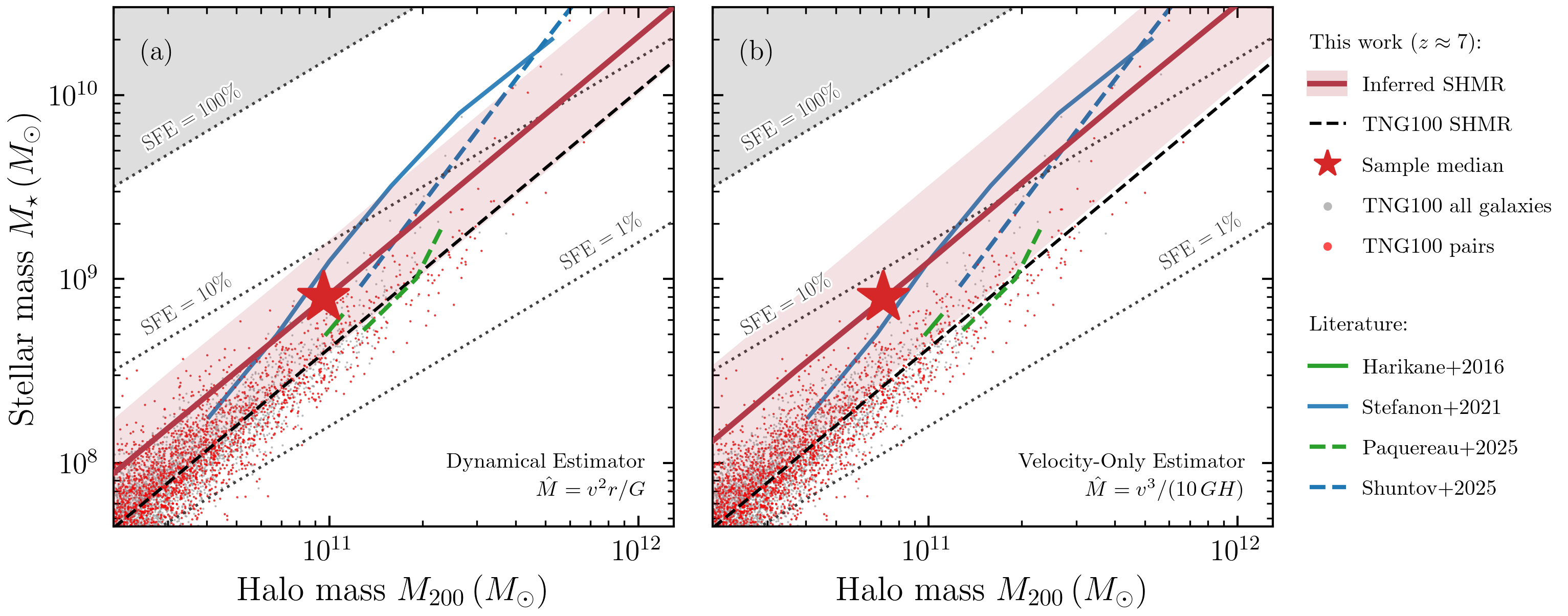}
    \caption{Stellar-to-halo mass relation at $z\approx7$ inferred using: (a) the dynamical-mass estimator $\hat{M}=v^2r/G$, and (b) the velocity-only estimator $\hat{M}=v^3/(10\,GH)$. We adopt the dynamical-mass estimate in panel (a) as our fiducial result. The red line and shaded region show the inferred SHMR and its 16th--84th percentile uncertainty, and the red star marks the sample median. Gray and red points show TNG100 individual  galaxies and galaxy pairs, respectively, while the black dashed line shows the TNG100 SHMR. Galaxy pairs follow the same SHMR as individual galaxies. In our inference, we fix the logarithmic slope of the SHMR to the TNG100 value $\alpha=1.4$ (Section~\ref{sec:inference}). Dotted lines indicate constant integrated star formation efficiencies (SFE) of 1\%, 10\%, and 100\%. Literature constraints at similar redshifts are shown for comparison, with green curves indicating galaxy-clustering constraints and blue curves indicating abundance-matching results.} 
    \label{fig:shmr}
\end{figure*}

\begin{deluxetable}{lcccc}
\tablecaption{Inferred Mean Halo Mass  $\log\,(M_\mathrm{h, 0}/M_\odot)$
\label{tab:halo_mass}}
\tablehead{
\colhead{} &
\multicolumn{2}{c}{Dynamical-mass estimator} &
\multicolumn{2}{c}{Velocity-only estimator} \\
\cline{2-3}\cline{4-5}
\colhead{$\alpha$} &
\colhead{$M_\star$--$M_{\rm h}$} &
\colhead{$L_{\rm UV}$--$M_{\rm h}$} &
\colhead{$M_\star$--$M_{\rm h}$} &
\colhead{$L_{\rm UV}$--$M_{\rm h}$}
}
\startdata
1.0 & $11.08^{+0.24}_{-0.22}$ & $11.04^{+0.20}_{-0.19}$ & $10.96^{+0.33}_{-0.32}$ & $10.89^{+0.30}_{-0.27}$ \\
1.4$^\dagger$ & $11.06^{+0.21}_{-0.20}$ & $11.03^{+0.21}_{-0.19}$ & $10.93^{+0.31}_{-0.29}$ & $10.88^{+0.30}_{-0.27}$ \\
2.0 & $11.05^{+0.22}_{-0.20}$ & $11.03^{+0.21}_{-0.20}$ & $10.91^{+0.31}_{-0.29}$ & $10.88^{+0.30}_{-0.27}$ \\
\enddata
\tablenotetext{\dagger}{Fiducial SHMR slope adopted in this work.}
\tablecomments{
Entries give the posterior median and 16th--84th percentile range of $\log\,(M_\mathrm{h, 0}/M_\odot)$ at $\log\,(M_{\rm \star, 0}/M_\odot)=9$ for the SHMR ($M_\star$--$M_{\rm h}$) and at $M_{\rm UV}=-20.3$ for the UV luminosity--halo mass relation ($L_{\rm UV}$--$M_{\rm h}$). These reference points are chosen to be near the corresponding sample medians. The parameter $\alpha$ is the assumed logarithmic slope of the corresponding galaxy-halo relation. The inferred mean halo mass is insensitive to $\alpha$.}
\end{deluxetable}

\section{Results}
\label{sec:results}
\subsection{Halo-mass constraints}
Using the 10 galaxy-pair systems in JADES, we find that the two estimators  introduced in Section~\ref{sec:method} yield consistent mean halo masses at the reference stellar mass. At a reference stellar mass of $\log\,(M_{\star,0}/M_\odot)=9$, close to the sample median, the dynamical-mass estimator yields $\log\,(M_\mathrm{h,0}/M_\odot)=11.06^{+0.21}_{-0.20}$, while the velocity-only estimator gives $10.93^{+0.31}_{-0.29}$. Their agreement provides a useful internal check, as the two estimators rely on different physical assumptions and respond differently to pair separation and projection. We adopt the dynamical-mass estimate as our fiducial result because of its smaller uncertainty.

The mild difference between the two estimates is expected from small-sample statistics, as it depends on the particular realization of projection angles and pair properties. Our mock simulations show that 34\% of  random realizations yield a difference between the two estimators larger than the observed value, and that the typical difference drops below 0.03~dex when using 100 pairs.

Our result is not sensitive to the adopted SHMR slope $\alpha$ near the sample median stellar mass, as shown in Table~\ref{tab:halo_mass}. We fix $\alpha$ to the TNG100 value because the current sample is too small to constrain it, but it can be inferred jointly with a larger sample. In any case, $\alpha$ serves only to combine galaxies of different stellar mass, and our dynamical method itself does not depend on the SHMR.

\subsection{Stellar-to-halo mass relation and SFE}
Our results imply that, at fixed halo mass, the stellar mass is 0.3~dex higher and the absolute UV magnitude $M_{\rm UV}$ is 0.4~mag brighter than in TNG100, at the $\sim1\sigma$ level. This stellar-mass offset may indicate that galaxy formation is more efficient than TNG100 predicts. The integrated star formation efficiency, which measures the fraction of the baryonic mass associated with a halo that has been converted into stars, is given by
\begin{equation}
    \epsilon_{\star,\rm int}= M_\star/(f_bM_{200}),
\end{equation}
where $f_b=0.16$ is the cosmic baryon fraction (\hspace{-3pt}\citealt{Planck2020CMB}). We infer $\epsilon_{\star,\rm int}=5.4^{+3.2}_{-2.1}\%$ at $\log\,(M_\star/M_\odot)=9$, about twice the TNG100 value of $2.6\%$, although this difference could still arise from statistical fluctuations.  As an interesting coincidence, we note that TNG100 also systematically underpredicts the stellar mass function relative to the measurements of \citet{Weibel2024} and \citet{Harvey2025}, but the galaxy number density at $10^9\,M_\odot$ becomes consistent with observations if the stellar masses are shifted upward by 0.3~dex.

\subsection{Comparison with the literature}
We compare our measurement with previous halo-mass estimates from abundance matching and galaxy clustering at $z\approx7$ \citep{Harikane2016, Stefanon2021, Paquereau2025, Shuntov2025}, as shown in Figure~\ref{fig:shmr}. At fixed stellar mass, our inferred halo mass is slightly lower, although it is broadly consistent within $\sim1\sigma$. Several effects could contribute to this offset. First, it may simply reflect statistical fluctuations given our limited sample size. Second, stellar-mass estimates in the literature rely on different assumptions and are subject to different systematic uncertainties, which propagate into the inferred SHMR. Third, the SHMR may evolve across $z=6$--$8$ \citep[e.g.,][]{Shuntov2025}, and because the literature samples have different redshift distributions, the effective redshift of each measured SHMR differs from ours, making the comparison in Figure~\ref{fig:shmr} not strictly at the same redshift. Finally, halo-mass inference from abundance matching and clustering is sensitive to sample selection, while current analyses usually rely on photometric samples, since large spectroscopic surveys at high redshift are expensive. Such samples, however, can be contaminated and incomplete: low-redshift interlopers, such as extreme emission-line galaxies and Balmer-break galaxies, may enter the sample, whereas faint post-burst galaxies \citep[e.g.,][]{Endsley2025Burstiness, Simmonds2025} and massive NIRCam-dark galaxies \citep{Sun2026NIRCamDark} may be missed. These effects may bias such indirect halo-mass inferences.

Our inferred SHMR is broadly consistent with empirical galaxy-halo models. At $\log\,(M_{200}/M_\odot)=11$, our stellar-to-halo mass ratio agrees with the prediction of \citet{Tacchella2018}, after converting their results to a \citet{Chabrier2003} IMF. It lies $\sim$0.2~dex above \textsc{UniverseMachine} \citep{Behroozi2019} and further above \textsc{Emerge} \citep{Moster2018}. These offsets do not necessarily challenge the models. They were calibrated on pre-JWST observations that provided limited rest-frame optical constraints, and their predictions may be revised as JWST measurements are incorporated. Their predictions also depend on different assumptions about how galaxies grow within their halos, which are poorly constrained by observations at high redshift.

\section{Discussion}
\subsection{Systematic uncertainties}
We discuss several potential systematic uncertainties in our halo mass inference. First, for the dynamical-mass estimator, we find no systematic bias larger than 0.05~dex in TNG100 for all the pairs separated by 0.6--1.5\,$R_{200}$. As an independent check, we repeat the test on the higher-resolution simulation TNG50 and find similar accuracy. Second, we select pairs by their projected separations, which could alter the distribution of viewing angles and hence of projection factors. We quantify this effect with end-to-end mocks from TNG100 that apply the same projected-separation selection and match the stellar-mass distribution of our sample. We find a bias below 0.06~dex, which is negligible at our current precision. Third, the \oiii\ velocity may not perfectly trace the center-of-mass velocity of the galaxy due to outflows. However, outflows mainly affect the broad component of the line, while the narrow component remains close to the systemic velocity of the galaxy. Moreover, any additional velocity scatter would only increase the inferred dynamical mass. If such an effect is significant, the true halo mass would be lower, increasing the difference from TNG100 and from other halo-mass measurements. Fourth, our velocity errors are dominated by the systematics of the wavelength calibration (Appendix~\ref{sec:wave_cal}). We may have overestimated these systematics, because the two galaxies of a close pair are observed together and their wavelength offsets likely cancel in part. The impact on our results is nevertheless small: halving the velocity errors changes the inferred halo mass by less than 0.02~dex.

\subsection{Dependence on the TNG100 simulation}
Our dynamical mass estimators are motivated by the first-infall dynamics of galaxy pairs. As discussed in Section~\ref{sec:motivation}, the conversion of gravitational potential energy into kinetic energy during infall gives $v^2r/G\approx M_{200}$ when the pair separation is near $R_{200}$. The TNG100 simulation confirms this normalization to within 0.05~dex, so we apply no additional calibration factor to Equations~(\ref{eq:mass-vir-los}) and (\ref{eq:mass-dyn-los}). From TNG100 we adopt only the intrinsic scatter of the estimators, marginalizing over it to obtain the halo mass and its uncertainty.

At separations near $R_{200}$, the pair dynamics is governed by the gravity of the dark matter halo and is therefore expected to be insensitive to the treatment of star formation and feedback. Dark matter dynamics is set by the cosmology and is thus well constrained even at high redshift. We do adopt the SHMR slope $\alpha$ from TNG100, but the inferred mean halo mass is insensitive to this choice over the range tested (Table~\ref{tab:halo_mass}), and $\alpha$ enters only in the inference of the SHMR, while our dynamical method itself does not depend on it.

\subsection{Prospects for future observations}
Our precision is currently limited by the sample size. Nonetheless, enlarging the sample is feasible, because a substantial fraction (10--30\%) of high-redshift galaxies reside in pair systems \citep{Duan2025, Puskas2025,Puskas2025a, Duan2026a}. The major obstacle has been that NIRSpec surveys often choose not to target galaxy companions, because their spectra may overlap and contaminate the primary targets. NIRCam/WFSS does not have target conflicts, but with its lower sensitivity it may not detect the companion at sufficient SNR. The NIRSpec/IFU also observes all the galaxies in its field of view, but the field is smaller than the expected $R_{200}$ at $z=7$. It reaches only the closest companions, where the halo mass is measured less precisely. However, with the recent development of the dense-shutter spectroscopy strategy that can extract emission lines in overlapping spectra in NIRSpec grating observations \citep{DEugenio2026DarkHorse}, a large sample of galaxy pairs can now be surveyed efficiently.

Our mock observations based on TNG100 predict that 50 galaxy pairs could constrain the SHMR halo-mass normalization to better than 0.08~dex. With 100 pairs, the normalization, slope, and intrinsic scatter of the relation could be jointly constrained to 0.08~dex, 0.3, and 0.1~dex, respectively. Because the slope probes the stellar-mass dependence of SFE and the scatter encodes the stochasticity of galaxy formation, these measurements would place strong constraints on models of rapid galaxy assembly in the early Universe.

The method can also be extended to higher redshift, although the choice of tracer becomes more restricted. NIRSpec gratings cover \oiii\ only out to $z\approx9$, beyond which the line falls outside their wavelength range. The [O\,{\footnotesize II}]\,$\lambda3729$ line remains accessible, but it is usually faint in these metal-poor galaxies. MIRI medium-resolution spectroscopy covers \oiii\ at $z\gtrsim10$ with a field of view comparable to $R_{200}$, albeit with lower sensitivity. The Atacama Large Millimeter/submillimeter Array (ALMA) can instead detect the [O\,{\footnotesize III}]\,88\,$\mu$m line \citep[e.g.,][]{Witstok2025gsz11}, which may offer an efficient way to survey pair velocities at $z>10$.

\section{Summary}
\label{sec:summary}
This work provides dynamical measurements of halo masses at $z\approx7$ using galaxy pairs.  We present two estimators (Equations~\ref{eq:mass-vir-los} and \ref{eq:mass-dyn-los}) that are accurate over $0.6<r/R_{200}<1.5$, with a mean bias below 0.05~dex, while remaining applicable at $r\lesssim 0.6\,R_{200}$ and $r\gtrsim 1.5\,R_{200}$, respectively, albeit with larger scatter. We validate the estimators with the TNG100 simulation, and recover the input halo mass in mock observations. 

Applying to 10 systems in the JADES survey, we infer a mean halo mass of $\log\,(M_{200}/M_\odot)=11.06\pm0.21$ at  $\log\,(M_{\star}/M_\odot)=9$, implying an integrated star formation efficiency of $5^{+3}_{-2}\%$. The inferred SHMR has a stellar mass 0.3~dex higher than in TNG100 at fixed halo mass. The current significance of the discrepancy is $1.3\sigma$ but could increase to $3\sigma$ with 50 pairs. Our halo mass is broadly consistent with previous inferences from abundance matching and galaxy clustering, although it is slightly lower than some of these estimates.

Our dynamical method is especially valuable at cosmic dawn, when indirect halo-mass inferences grow increasingly uncertain as the galaxy-halo connection becomes more stochastic. Galaxy pair dynamics, in contrast, directly probes the halo gravitational potential. Pairs at these epochs offer a further advantage: they are often still in the early stages of infall, before substantial orbital evolution and phase mixing, so that the intrinsic scatter of our estimators is small.

Future observations of larger galaxy-pair samples will place stringent constraints on the normalization, slope, scatter, and evolution of the SHMR. Enlarging the sample is feasible, because 10--30\% of galaxies at $z=3$--$9$ are expected to reside in pair systems. These measurements will anchor the galaxy-halo connection and shed light on the efficiency and stochasticity of galaxy formation in the early Universe.

\begin{acknowledgements}
We thank Yueying Ni and Anna de Graaff for insightful discussions. This work is based on observations made with the NASA/ESA/CSA James Webb Space Telescope. The data were obtained from the Mikulski Archive for Space Telescopes at the Space Telescope Science Institute, which is operated by the Association of Universities for Research in Astronomy, Inc., under NASA contract NAS 5-03127 for JWST. Support for program \#3215 was provided by NASA through a grant from the Space Telescope Science Institute, which is operated by the Association of Universities for Research in Astronomy, Inc., under NASA contract NAS 5-03127. We thank the CONGRESS and FRESCO teams for developing their observing programs with zero-exclusive-access periods. This research made use of the lux supercomputer at UC Santa Cruz which is funded by NSF MRI grant AST 1828315.  

Z.W. and D.J.E. are supported by the Simons Foundation Investigator program. D.J.E., B.D.J., B.E.R.,  J.M.H., Z.J., and Y.Z. acknowledge support from the NIRCam Science Team contract to the University of Arizona, NAS5-02105. D.J.E. is also supported by NASA through a grant from the Space Telescope Science Institute, which is operated by the Association of Universities for Research in Astronomy, Inc., under NASA contract NAS5-03127. S.T. acknowledges support by the Royal Society Research Grant G125142. B.E.R. also acknowledges support from JWST Program 3215. S.A. acknowledges grant PID2021-127718NB-I00 funded by the Spanish Ministry of Science and Innovation/State Agency of Research (MICIN/AEI/ 10.13039/501100011033). A.J.B. acknowledges funding from the ``FirstGalaxies'' Advanced Grant from the European Research Council (ERC) under the European Union's Horizon 2020 research and innovation program (Grant agreement No. 789056). Q.D. acknowledges support from a PhD studentship awarded by Trinity College, University of Cambridge. J.M.H. acknowledges support from the Evolving Universe Fellowship, which is made possible by a generous donation from Dr. Keiko Miwa Ross; J.M.H. also acknowledges support from JWST Program 8544.  D.P. acknowledges support by the Science and Technology Facilities Council (STFC), by the ERC through Advanced Grant 695671 ``QUENCH'', and by the UKRI Frontier Research grant RISEandFALL. D.P. also acknowledges support by the Huo Family Foundation through a P.C. Ho PhD Studentship. P.R. acknowledges support from the University of Texas at Austin Cosmic Frontier Center. H.U. acknowledges support by the Max Planck Society through the Lise Meitner Excellence Program. H.U. acknowledges funding by the European Union (ERC APEX, 101164796). Views and opinions expressed are however those of the authors only and do not necessarily reflect those of the European Union or the European Research Council Executive Agency. Neither the European Union nor the granting authority can be held responsible for them.

We acknowledge the use of Claude Code (Anthropic) and Codex (OpenAI) for assistance in developing the analysis code and for suggesting improvements to the clarity and flow of the text. The authors have reviewed all content and take full responsibility for the scientific results and conclusions.
\end{acknowledgements}

\software{
\texttt{Astropy} (\hspace{-4pt}\citealt{astropy2013, astropy2018, astropy2022ApJ}),
\texttt{dynesty} \citep{Speagle2020},
JWST Science Calibration Pipeline \citep{Bushouse2023jwst},
\texttt{Prospector} \citep{Johnson2021ApJS}.
}

\appendix
\twocolumngrid
\restartappendixnumbering

\section{Wavelength Calibration}
\label{sec:wave_cal}
Here we quantify the residual wavelength-calibration systematics of each instrument relative to the high-resolution grating G395H. We select all galaxies at $z=6$--$8$ in the JADES NIRSpec catalog that are observed with more than one instrument, and measure their redshifts from the \oiii\ line with the same fitting pipeline used for the pair sample (Section~\ref{sec:sample}). The fit accounts for noise correlations between wavelength channels, so that the resulting uncertainties appropriately reflect the random noise. We further restrict the comparison to relatively bright sources with line SNR $>5$. We do not distinguish between the NIRCam/WFSS filters, because the dispersion is provided by the same grism. For each galaxy, we then compute the velocity offset of each instrument relative to G395H, and apply sigma clipping that removes $5\sigma$ outliers.

Inferring the systematics is not trivial, because they must be distinguished from the random measurement errors. The random errors vary from galaxy to galaxy with the line SNR and are sometimes larger than the systematics. We therefore adopt a Bayesian framework that infers the systematic offset and scatter simultaneously. We formalize the problem as follows: given velocity offsets $v_i$ with measurement errors $v_{{\rm err},i}$, we infer the intrinsic mean $\mu$ and scatter $\sigma$, assuming that the offset of the $i$-th galaxy is drawn from a Gaussian distribution,

\begin{equation}
v_i \sim \mathcal N(\mu,\; v_{{\rm err},i}^2+\sigma^2).
\end{equation}

\noindent
The log-posterior probability is

\begin{equation}
\begin{aligned}
\log\, p(\mu,\,\sigma\mid v)
= &-\frac{1}{2}\sum_i \bigg[
    \log\left(v_{{\rm err},i}^2+\sigma^2\right)
    + \frac{(v_i-\mu)^2}{v_{{\rm err},i}^2+\sigma^2}
\bigg]\\
&+ \log\,p(\mu) + \log\,p(\sigma)  + \mathrm{const.},
\end{aligned}
\end{equation}

\noindent
where we adopt a uniform prior for $\mu$ and the Jeffreys prior on $\sigma$,

\begin{equation}
    p(\sigma) \propto \sigma \left[ \sum_i \frac{1}{(v_{{\rm err},i}^2+\sigma^2)^2} \right]^{1/2}.
\end{equation}

\noindent
The Jeffreys prior is non-informative and reduces to $p(\sigma)\propto 1/\sigma$ when the measurement errors are negligible. We sample the posterior on a grid covering $-50<\mu<50$\kms\ and $0<\sigma<50$\kms\ with a step of 0.2\kms, and adopt the maximum {\it a posteriori} (MAP) values as our estimates. The results are listed in Table~\ref{tab:systematics}. The half-widths of the posterior 16th--84th percentile intervals of both $\mu$ and $\sigma$ are $\sim$2, 2, and 6\kms\ for G235M, G395M, and NIRCam/WFSS, respectively. As a consistency check, we also estimate the bias and scatter directly from the median and standard deviation of the velocity offsets, and find agreement with the Bayesian values to within 20\%.

Our results suggest that the wavelength measurements from G235M are consistent with those from G395H. However, the other two instruments, G395M and NIRCam/WFSS, show noticeable systematic biases and scatters. Offsets of similar magnitude, in units of NIRSpec pixels, have also been reported between the prism and the gratings \citep{Bunker2024JADES, scholtz_jades_2026}. We correct for the biases and propagate the systematic scatters into our velocity errors. 

For close pairs, however, the uncertainty in the relative velocity may be smaller than the systematic scatter inferred from the full sample. When two nearby sources are observed at the same time, their wavelength offsets are likely correlated, shifting in the same direction and partially canceling in the difference. Our error budget is therefore conservative. The measured velocities themselves are unaffected, since both galaxies of a pair are corrected in the same way.

\begin{deluxetable}{cccc}
\setlength{\tabcolsep}{10pt}
\vspace{10pt}
\tablecaption{Instrumental Velocity Systematics (\kms)}
\tablehead{
\colhead{Instrument} & \colhead{G235M} &  \colhead{G395M} & \colhead{NIRCam/WFSS}
}
\startdata
Bias  & $-1$ & $-25$ & $15$ \\
Scatter & 7 & $12$ & 21 \\
\enddata
\tablecomments{Systematics are estimated relative to the NIRSpec high-resolution grating (G395H), using redshifts derived from the [O\,{\tiny III}]\,$\lambda5008$ emission line. The comparison samples comprise 14, 32, and 8 compact, high-SNR galaxies at $z=6$--$8$ (not restricted to the pair sample) whose G395H observations overlap with G235M, G395M, or NIRCam/WFSS, respectively.}
\label{tab:systematics}
\end{deluxetable}

\section{Galaxy Properties}

Table~\ref{tab:detail} lists the properties of the individual galaxies in the 10 systems of Table~\ref{tab:pair_sample}. For each galaxy we list its JADES NIRCam catalog ID, coordinates, \oiii\ redshift and line SNR from every instrument in which it is detected, rest-frame UV absolute magnitude, stellar mass, and NIRCam photometry. 

\begin{deluxetable*}{rrrrrrrrrrrrr}
\tablecaption{Properties of the individual galaxies in the galaxy pair sample. \label{tab:detail}}
\tablehead{\colhead{ID} & \colhead{System} & \colhead{R.A. (deg)} & \colhead{Decl. (deg)} & \colhead{$z_{\rm [O\,III]}$} & \colhead{SNR} & \colhead{Instrument} & \colhead{Program} & \colhead{$M_{\rm UV}$} & \colhead{$\log(M_\star/M_\odot)$} & \colhead{F115W (nJy)} & \colhead{F277W (nJy)} & \colhead{F444W (nJy)}}
\startdata
32002 & Pair3 & 53.091650 & $-27.876905$ & $6.1644 \pm 0.0003$ & 8.9 & G235M & JADES & $-17.75 \pm 0.25$ & $7.53^{+0.30}_{-0.20}$ & $8.86 \pm 2.30$ & $11.79 \pm 1.68$ & $9.56 \pm 1.88$ \\
 &  &  &  & $6.1644 \pm 0.0006$ & 7.6 & G395M & JADES &  &  &  &  &  \\
68395 & Pair2 & 53.069992 & $-27.852705$ & $6.1266 \pm 0.0007$ & 3.6 & F356W & JADES & $-17.47 \pm 0.17$ & $7.72^{+0.24}_{-0.22}$ & $6.89 \pm 1.15$ & $9.99 \pm 0.94$ & $8.37 \pm 1.09$ \\
 &  &  &  & $6.1265 \pm 0.0011$ & 4.8 & F322W2 & JADES &  &  &  &  &  \\
68872 & Pair2 & 53.070100 & $-27.852276$ & $6.1261 \pm 0.0005$ & 16.6 & F356W & JADES & $-19.59 \pm 0.03$ & $8.55^{+0.07}_{-0.08}$ & $48.58 \pm 1.18$ & $48.75 \pm 0.97$ & $52.44 \pm 1.08$ \\
 &  &  &  & $6.1270 \pm 0.0005$ & 13.5 & F322W2 & JADES &  &  &  &  &  \\
154775 & Pair9 & 53.169567 & $-27.738060$ & $7.2423 \pm 0.0006$ & 12.2 & F444W & FRESCO & $-20.27 \pm 0.03$ & $8.37^{+0.16}_{-0.10}$ & $71.30 \pm 2.00$ & $62.26 \pm 1.28$ & $115.02 \pm 1.39$ \\
169896 & Pair3 & 53.091164 & $-27.877272$ & $6.1656 \pm 0.0002$ & 8.5 & G395H & JADES & $-18.49 \pm 0.14$ & $8.11^{+0.25}_{-0.21}$ & $17.42 \pm 2.31$ & $23.61 \pm 1.66$ & $26.49 \pm 1.87$ \\
 &  &  &  & $6.1664 \pm 0.0003$ & 6.5 & G235M & JADES &  &  &  &  &  \\
 &  &  &  & $6.1661 \pm 0.0006$ & 7.5 & G395M & JADES &  &  &  &  &  \\
 &  &  &  & $6.1652 \pm 0.0007$ & 5.5 & F356W & JADES &  &  &  &  &  \\
 &  &  &  & $6.1653 \pm 0.0009$ & 4.6 & F322W2 & JADES &  &  &  &  &  \\
175729 & Pair10 & 53.060580 & $-27.866024$ & $7.8823 \pm 0.0001$ & 21.1 & G395H & JADES & $-19.25 \pm 0.12$ & $8.21^{+0.29}_{-0.08}$ & $24.74 \pm 2.80$ & $33.12 \pm 1.91$ & $64.93 \pm 2.29$ \\
 &  &  &  & $7.8825 \pm 0.0004$ & 40.9 & G395M & JADES &  &  &  &  &  \\
 &  &  &  & $7.8809 \pm 0.0008$ & 8.3 & F444W & JADES &  &  &  &  &  \\
175837 & Pair10 & 53.060210 & $-27.865717$ & $7.8834 \pm 0.0009$ & 4.9 & F444W & JADES & $-18.05 \pm 0.32$ & $8.49^{+0.32}_{-0.23}$ & $8.22 \pm 2.86$ & $25.70 \pm 1.82$ & $54.29 \pm 2.26$ \\
187025 & Pair7 & 53.106040 & $-27.848225$ & $6.9069 \pm 0.0006$ & 55.1 & F444W & JADES & $-20.60 \pm 0.04$ & $8.64^{+0.06}_{-0.05}$ & $103.22 \pm 4.13$ & $94.83 \pm 1.91$ & $232.58 \pm 2.20$ \\
 &  &  &  & $6.9063 \pm 0.0006$ & 51.4 & F356W & JADES &  &  &  &  &  \\
 &  &  &  & $6.9068 \pm 0.0006$ & 64.1 & F322W2 & JADES &  &  &  &  &  \\
187028 & Pair7 & 53.106174 & $-27.848020$ & $6.9022 \pm 0.0007$ & 5.8 & F356W & JADES & $-18.73 \pm 0.22$ & $8.14^{+0.32}_{-0.16}$ & $18.49 \pm 4.09$ & $28.21 \pm 1.85$ & $43.00 \pm 2.14$ \\
 &  &  &  & $6.9036 \pm 0.0007$ & 5.2 & F444W & JADES &  &  &  &  &  \\
 &  &  &  & $6.9028 \pm 0.0007$ & 6.4 & F322W2 & JADES &  &  &  &  &  \\
187193 & Pair7 & 53.105286 & $-27.847904$ & $6.9054 \pm 0.0008$ & 6.3 & F444W & JADES & $-19.62 \pm 0.10$ & $8.35^{+0.19}_{-0.24}$ & $42.19 \pm 4.08$ & $37.91 \pm 1.88$ & $61.12 \pm 2.17$ \\
 &  &  &  & $6.9050 \pm 0.0009$ & 4.6 & F322W2 & JADES &  &  &  &  &  \\
219057 & Pair9 & 53.168686 & $-27.737202$ & $7.2406 \pm 0.0007$ & 7.9 & F444W & FRESCO & $-19.03 \pm 0.09$ & $7.93^{+0.09}_{-0.09}$ & $22.84 \pm 1.92$ & $25.34 \pm 1.21$ & $51.84 \pm 1.30$ \\
1000741 & Pair8 & 189.074100 & $62.222305$ & $7.0344 \pm 0.0008$ & 4.6 & F444W & FRESCO & $-18.92 \pm 0.19$ & $7.95^{+0.17}_{-0.13}$ & $21.45 \pm 4.17$ & $23.16 \pm 2.75$ & $34.40 \pm 3.53$ \\
1000905 & Pair8 & 189.073680 & $62.223460$ & $7.0354 \pm 0.0008$ & 8.7 & F444W & FRESCO & $-20.21 \pm 0.06$ & $9.26^{+0.12}_{-0.25}$ & $70.30 \pm 4.17$ & $73.34 \pm 2.44$ & $118.09 \pm 3.59$ \\
1003743 & Pair1 & 189.077870 & $62.235966$ & $6.0510 \pm 0.0006$ & 10.9 & F356W & CONGRESS & $-19.45 \pm 0.14$ & $8.14^{+0.35}_{-0.20}$ & $43.34 \pm 5.95$ & $44.99 \pm 3.01$ & $49.94 \pm 4.28$ \\
1003962 & Pair1 & 189.078310 & $62.236780$ & $6.0470 \pm 0.0005$ & 18.6 & F356W & CONGRESS & $-19.26 \pm 0.15$ & $8.24^{+0.22}_{-0.18}$ & $36.64 \pm 5.38$ & $36.68 \pm 3.08$ & $46.42 \pm 4.18$ \\
1006478 & Pair4 & 189.135240 & $62.244667$ & $6.7223 \pm 0.0006$ & 22.0 & F356W & CONGRESS & $-18.43 \pm 0.23$ & $9.21^{+0.15}_{-0.14}$ & $14.62 \pm 3.45$ & $78.99 \pm 3.17$ & $137.63 \pm 3.83$ \\
1006564 & Pair4 & 189.135650 & $62.244854$ & $6.7299 \pm 0.0006$ & 13.6 & F356W & CONGRESS & $-18.55 \pm 0.21$ & $8.04^{+0.20}_{-0.16}$ & $16.27 \pm 3.42$ & $17.21 \pm 3.01$ & $15.04 \pm 3.77$ \\
1010806 & Pair6 & 189.153990 & $62.259544$ & $6.7596 \pm 0.0009$ & 4.0 & F356W & CONGRESS & $-18.61 \pm 0.23$ & $7.78^{+0.22}_{-0.17}$ & $17.13 \pm 4.07$ & $16.32 \pm 3.93$ & $14.45 \pm 4.19$ \\
1010816 & Pair6 & 189.151950 & $62.259644$ & $6.7606 \pm 0.0006$ & 27.3 & F356W & CONGRESS & $-19.81 \pm 0.09$ & $9.89^{+0.11}_{-0.13}$ & $51.79 \pm 4.43$ & $82.25 \pm 3.98$ & $301.94 \pm 4.76$ \\
 &  &  &  & $6.7597 \pm 0.0006$ & 20.1 & F444W & FRESCO &  &  &  &  &  \\
1015068 & Pair5 & 189.138530 & $62.275610$ & $6.7451 \pm 0.0007$ & 9.2 & F356W & CONGRESS & $-20.03 \pm 0.05$ & $9.33^{+0.12}_{-0.13}$ & $63.30 \pm 2.99$ & $105.77 \pm 3.01$ & $124.84 \pm 3.62$ \\
1027155 & Pair4 & 189.133680 & $62.244610$ & $6.7208 \pm 0.0006$ & 18.1 & F356W & CONGRESS & $-19.72 \pm 0.08$ & $8.29^{+0.17}_{-0.14}$ & $48.10 \pm 3.51$ & $49.79 \pm 3.21$ & $48.87 \pm 3.81$ \\
1031111 & Pair5 & 189.138080 & $62.274440$ & $6.7486 \pm 0.0006$ & 26.6 & F356W & CONGRESS & $-19.37 \pm 0.10$ & $8.60^{+0.20}_{-0.16}$ & $34.33 \pm 3.15$ & $59.04 \pm 3.26$ & $65.75 \pm 3.70$ \\
 &  &  &  & $6.7480 \pm 0.0007$ & 6.1 & F444W & FRESCO &  &  &  &  &  \\
1031196 & Pair5 & 189.139140 & $62.275314$ & $6.7452 \pm 0.0006$ & 11.7 & F356W & CONGRESS & $-19.64 \pm 0.07$ & $9.14^{+0.27}_{-0.25}$ & $44.37 \pm 3.01$ & $87.77 \pm 3.02$ & $112.09 \pm 3.58$
\enddata
\tablecomments{Galaxies are ordered by NIRCam ID. The System column refers to Table~\ref{tab:pair_sample}. Coordinates, $M_{\rm UV}$, $M_\star$, and photometry are listed once per galaxy; any rows below give the redshift and [O\,{\tiny III}]\,$\lambda5008$ SNR of the same galaxy measured with a different instrument. G395H, G235M, and G395M are NIRSpec gratings. F322W2, F356W, and F444W are NIRCam/WFSS filters. The first row of each galaxy gives its adopted redshift following the instrument  preference discussed in Section~\ref{sec:sample}. Redshifts are measured by fitting the [O\,{\tiny III}]\,$\lambda5008$ line as described in Section~\ref{sec:sample}, and are listed after correction for the instrumental systematics in Table~\ref{tab:systematics}.  Stellar masses are from the JADES Data Release 5 catalog \citep{Duan2026}.}
\end{deluxetable*}

\bibliography{reference}{}
\bibliographystyle{aasjournalv7}

\end{document}